\documentclass{ws-rv9x6}
\usepackage{subfigure}   
\usepackage{ws-rv-thm}   
\usepackage{ws-rv-van}   
\makeindex
\def\CA {{\cal A}}

\newcommand{\be}{\begin{equation}}
\newcommand{\ee}{\end{equation}}

\newcommand{\beqa}{\begin{eqnarray}}
\newcommand{\eeqa}{\end{eqnarray}}
\newcommand{\nn}{\nonumber}

\begin{document}

\chapter[Using World Scientific's Review Volume Document Style]{Lars Brink  \\    Colleague, Friend and Collaborator\\  
}
\author[G.Savvidy]{George Savvidy 
}

\address{INPP, NCSR Demokritos, Agia Paraskevi, Athens, Greece \\
savvidy@inp.demokritos.gr 
}

\begin{abstract}
The Landau-Nordita conference that was held in Moscow in 1981 was the first place where I met Lars.  This  "marvellous meeting", as Lars quoted in one of his reminiscence article, was organised by Alan Luther. It was an event that generated a long-lasting friendship of scientists from East and West, provided a background for the future collaborations and, in particular, my own collaboration with Lars on a supersymmetric high-spin extension of the Poincar\'e group. \\

 \end{abstract}


\body


\section{\it SUSY extension of high spin Poincar\'e Algebra \cite{Antoniadis:2011re}}

The collaboration with Lars was not only a great privilege but also a joyful friendship. The algebra we were interested in was a supersymmetric extension of the high-spin Poincar\'e algebra introduced earlier in \cite{Savvidy:2008zy, Savvidy:2010vb, Savvidy:2010kw} \footnote{The collaboration with Lars and Ignatios on the supersymmetric high spin extension of the Poincar\'e  algebra started in the fall of 2010.  At that period Lars was busy working in the Nobel Committee for Physics, taking part in the Pune conference in India and spent a month at the Institute of Advance Study in Princeton. } . The algebra (\ref{extensionofpoincarealgebra}) naturally appeared in the high-spin extension of the Yang Mills theory suggested in \cite{Savvidy:2005fi, Savvidy:2010vb, Savvidy:2015jgv} . Lars was interested in finding out if there existed a supersymmetric extension of the tensor gauge fields Lagrangian proposed in \cite{Savvidy:2005fi, Savvidy:2010vb, Savvidy:2015jgv} or the supersymmetric extension of high-spin Poincar\'e algebra (\ref{extensionofpoincarealgebra}) \cite{Savvidy:2008zy, Savvidy:2010vb, Savvidy:2010kw} .  The  non-Abelian tensor gauge fields  were defined as rank-$(s+1)$ tensors
\cite{Savvidy:2005fi,Savvidy:2015jgv}
$$
A^{a}_{\mu\lambda_1 ... \lambda_{s}}(x),~~~~~s=0,1,2,...
$$
and were totally symmetric with respect to the indices $  \lambda_1 ... \lambda_{s}  $ and had no symmetries with
respect to the first index  $\mu$. The index $a$ numerates the generators $L^a$ of the Lie algebra  of a compact Lie group G. These tensor fields appear in the expansion of the gauge field $\CA_{\mu}(x,e)$ over the unit  polarisation vector
$e_{\lambda}$  \cite{Savvidy:2006yj, Savvidy:2006yk} :
\be\label{gaugefield}
{\cal A}_{\mu}(x,e)=\sum_{s=0}^{\infty} {1\over s!} ~A^{a}_{\mu\lambda_{1}...
\lambda_{s}}(x)~L^{a}e_{\lambda_{1}}...e_{\lambda_{s}}, 
\ee
where $L^{a}_{\lambda_1 ... \lambda_{s}} = L^a e_{\lambda_1}...e_{\lambda_s}$  are the "gauge generators" of the following current algebra  \cite{Savvidy:2008zy, Savvidy:2010vb, Savvidy:2010kw}
\be\label{curralg}
[L^{a}_{\lambda_1 ... \lambda_{i}}, L^{b}_{\lambda_{i+1} ... \lambda_{s}}]=if^{abc}
L^{c}_{\lambda_1 ... \lambda_{s} },~~~~~s=0,1,2,...
\ee
This current  algebra has infinitely many "gauge generators" $L^{a}_{\lambda_1 ... \lambda_{s}}$, they are gauge generators because they carry the isospin and Lorentz indices. The  generators $L^{a}_{\lambda_1 ... \lambda_{s}}$ are symmetric space-time tensors and  the full algebra should includes the Poincar\'e generators $P_{\mu},~M_{\mu\nu}$.  It has the following form \cite{Savvidy:2008zy, Savvidy:2010vb, Savvidy:2010kw} :
\beqa\label{extensionofpoincarealgebra}
~&&[P^{\mu},~P^{\nu}]=0,\nn\\
~&&[M^{\mu\nu},~P^{\lambda}] = i(\eta^{\lambda \nu}~P^{\mu}
- \eta^{\lambda \mu }~P^{\nu}) ,\nn\\
~&&[M^{\mu \nu}, ~ M^{\lambda \rho}] = i(\eta^{\mu \rho}~M^{\nu \lambda}
-\eta^{\mu \lambda}~M^{\nu \rho} +
\eta^{\nu \lambda}~M^{\mu \rho}  -
\eta^{\nu \rho}~M^{\mu \lambda} ),\nonumber\\
~&&[P^{\mu},~L_{a}^{\lambda_1 ... \lambda_{s}}]=0, \nn\\
~&&[M^{\mu \nu}, ~ L_{a}^{\lambda_1 ... \lambda_{s}}] = i(
\eta^{\lambda_1\nu } L_{a}^{\mu \lambda_2... \lambda_{s}}
-\eta^{\lambda_1\mu} L_{a}^{\nu\lambda_2... \lambda_{s}}
+...+
\eta^{\lambda_s\nu } L_{a}^{\lambda_1... \lambda_{s-1}\mu } -
\eta^{\lambda_s\mu } L_{a}^{\lambda_1... \lambda_{s-1}\nu } ),\nonumber\\
~&&[L_{a}^{\lambda_1 ... \lambda_{i}}, L_{b}^{\lambda_{i+1} ... \lambda_{s}}]=if_{abc}~
L_{c}^{\lambda_1 ... \lambda_{s} } ,     ~~~(\mu,\nu,\rho,\lambda=0,1,2,3; ~~~~~s=0,1,2,... )
\eeqa
It is an extension of the Poincar\'e algebra by "gauge generators" $L^{a}_{\lambda_1 ... \lambda_{s}}$, which are the elements of the current algebra (\ref{curralg}).  

In our first attempt to extend the algebra (\ref{extensionofpoincarealgebra}) to a supersymmetric case we were trying to consider an infinite set of spinor-tensor generators $Q^{i}_{\alpha \lambda_1 ... \lambda_{s}} \leftrightarrow L^{a}_{\lambda_1 ... \lambda_{s} }$, but this did not work\footnote{Lars' message on February 15 2011. "I like it that you changed the susy generators to $Q$. In fact the extended susy's seem to be a simple extension. It is interesting that one cannot extend the Lie algebra to fermionic generators but only the Poincar\'e ones. If you have massless fields you really have a conformal symmetry. There are no continuous reps in the conformal group as far as I know. What happens with the conformal symmetry if you have all the tensor fields?" \label{fut1}}. Finally during the conference in Florida the following extension was suggested:
\beqa\label{supergaugePoincare}
~&&[P^{\mu},~P^{\nu}]=0, \label{extensionofpoincarealgebra1}\\
~&&[M^{\mu\nu},~P^{\lambda}] = i(\eta^{\lambda \nu}~P^{\mu}
- \eta^{\lambda \mu }~P^{\nu}) ,\nn\\
~&&[M^{\mu \nu}, ~ M^{\lambda \rho}] = i(\eta^{\mu \rho}~M^{\nu \lambda}
-\eta^{\mu \lambda}~M^{\nu \rho} +
\eta^{\nu \lambda}~M^{\mu \rho}  -
\eta^{\nu \rho}~M^{\mu \lambda} ),\nn\\
\nn\\
~&&[P^{\mu},~L_{a}^{\lambda_1 ... \lambda_{s}}]=0,  \\
~&&[P^{\mu},~Q^{i}_{\alpha}]=0, \nn\\
~&&[M^{\mu \nu}, ~ L_{a}^{\lambda_1 ... \lambda_{s}}] = i(
\eta^{\lambda_1\nu } L_{a}^{\mu \lambda_2... \lambda_{s}}
-\eta^{\lambda_1\mu} L_{a}^{\nu\lambda_2... \lambda_{s}}
+...+
\eta^{\lambda_s\nu } L_{a}^{\lambda_1... \lambda_{s-1}\mu } -
\eta^{\lambda_s\mu } L_{a}^{\lambda_1... \lambda_{s-1}\nu } ),\nonumber\\
~&&[M^{\mu \nu}, ~ Q^{i}_{\alpha}] = {i\over 2} (
\gamma^{\mu\nu} Q^{i})_{\alpha} ,~~~~~~
\gamma^{\mu\nu}={1\over 2} [\gamma^{\mu},\gamma^{\nu}]  \nn\\
\nn\\
\label{alakacss}
~&&[L_{a}^{\lambda_1 ... \lambda_{n}}, L_{b}^{\lambda_{n+1} ... \lambda_{s}}]=if_{abc}
L_{c}^{\lambda_1 ... \lambda_{s} }     ,  ~~~  s=0,1,2,...   \nn\\
~&&\{ Q^{i}_{\alpha},~ Q^{j}_{\beta} \}=-2~\delta^{ij} (\gamma^{\mu} C)_{\alpha\beta}
 P_{\mu},~~~~~i=1,...,N  \nn\\
~&&[L_{a}^{\lambda_1 ... \lambda_{s}}, Q^{i}_{\alpha} ]=0~.
\eeqa
The supersymmetric generalisations of high spin de Sitter and conformal groups  were also suggested in \cite{Antoniadis:2011re} . These algebras were defined  in our publication in the Journal of Mathematical Physics \cite{Antoniadis:2011re} . The publication was the result of a long-lasting friendship that started during the Landau-Nordita  meeting that was held in Moscow in 1981. Below are the recollections of Landau-Nordita  meeting and the exchange of messages with Lars during our collaboration on these algebras\footnote{The summary of scientific achievements of Lars Brink can be found in the article of  Bengt E.W. Nilsson and Bj\"orn Jonson  \cite{Nilsson:2024fmi} .}.  
 \begin{figure}
\centerline{\includegraphics[width=5cm]{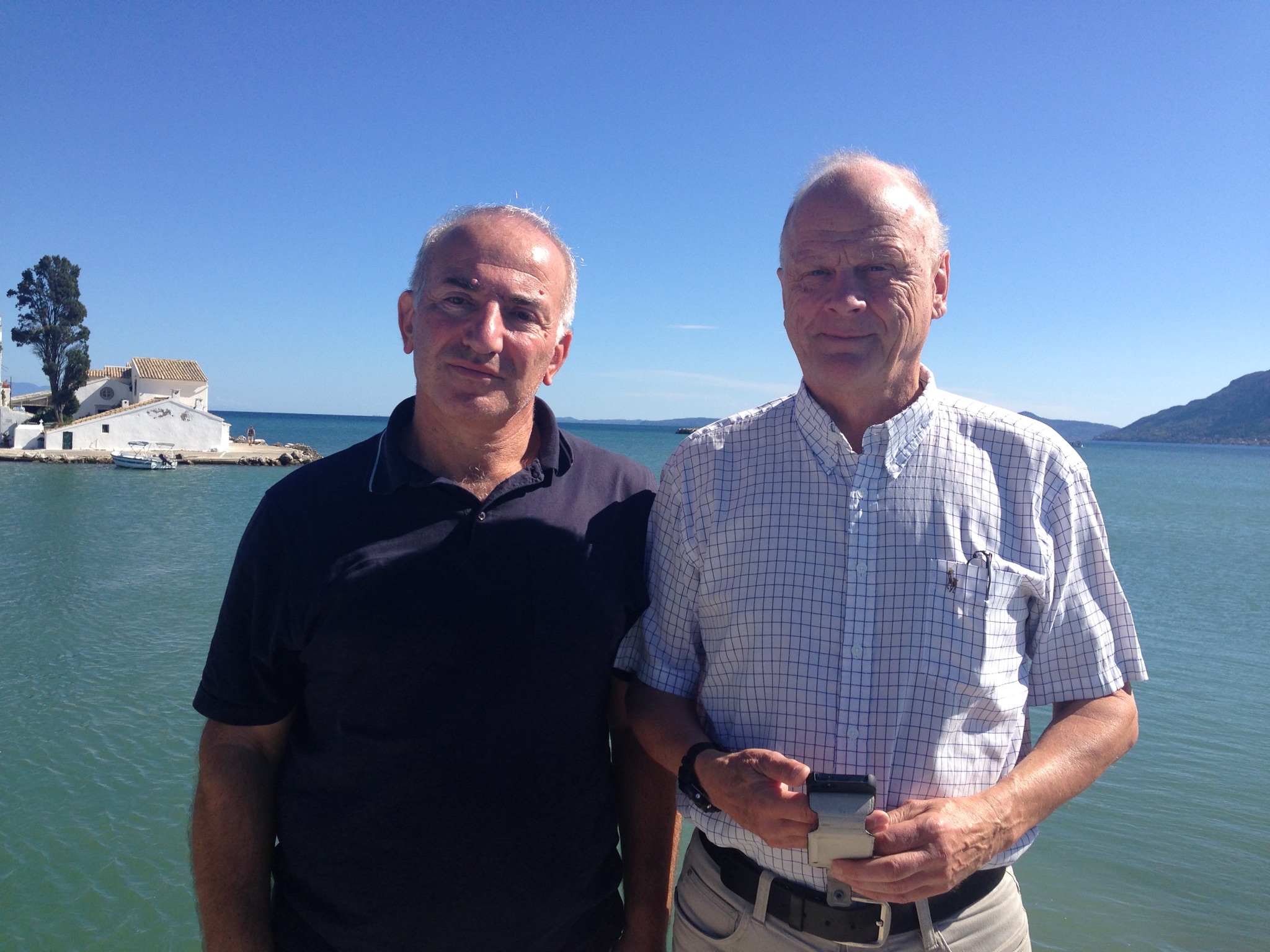}}
\caption{ During the Corfu meeting in 2013 Lars presented the overview of the discovery of  Higgs boson and the Nobel prise laureates François Englert and Peter W. Higgs.  The photo with Lars during the Corfu meeting, 2013.} 
\label{fig1}
\end{figure}

\section{\it The Landau-Nordita  Meeting  in 1981}

\begin{figure}
\centerline{\includegraphics[width=5cm]{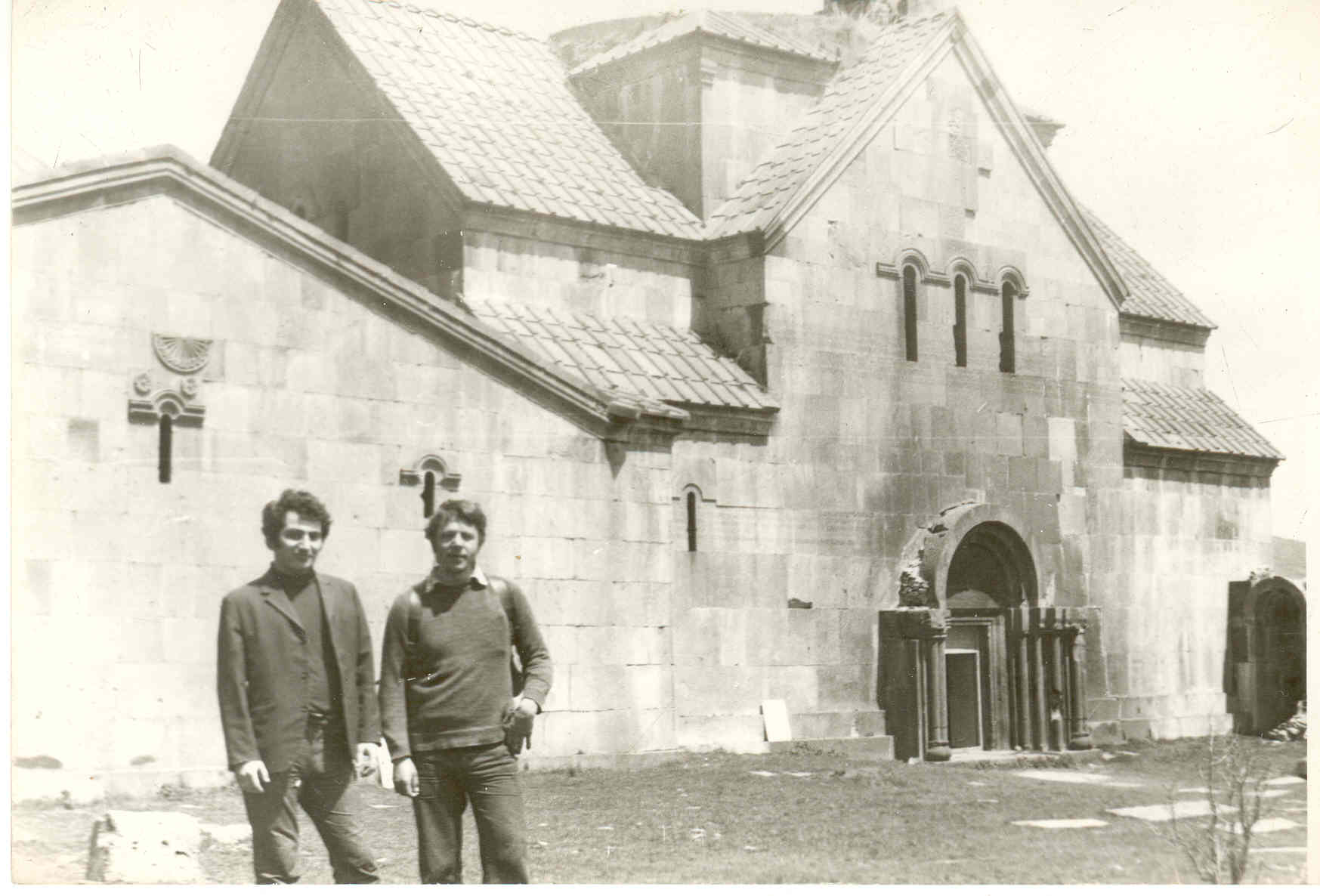}}
\caption{
The Nuclear Physics Section of the Academy of Sciences was held in Moscow in 1976 \cite{programe1976}. There was a small subsection (33-T)  \cite{programe1976} devoted to the Yang Mills theory with a few participants, including  Alexander Belavin and Alexander Migdal, both  from the Landau Institute. Many of the conference sections  were devoted to the S-matrix theory, Regge theory, and the phenomenology of strong interactions. I gave  a talk on the polarisation of the Yang Mills theory vacuum, the non-Abelian extension of Heisenberg-Euler Lagrangian in QED.  After the talk Belavin invited me to the Landau Institute in Cheronolovka. We spent the whole night discussing the polarisation of the YM theory vacuum presented earlier in my talk.  Alexander was interested in investigating the polarisation of the YM vacuum by instanton field \cite{Belavin:1976vj, tHooft:1976snw}.This event created a long-lasting friendship with the theory group of the Landau Institute and subsequent common meetings in Cheronolovka and Yerevan. Photo with Alexander Belavin in Armenia, 1978.} \label{fig2}
\end{figure}

I first met Lars Brink in 1981 during the Landau-Nordita  meeting that was held in Moscow\footnote{As Lars quoted in his article \cite{doi:10.1142/5025}, "The next time I came to the Soviet Union was in 1981 when there was the first Landau-Nordita meeting. It was a marvellous meeting with unmatched hospitality from the Russian side. Alan Luther who had organized the Nordita delegation, where only two or three people really worked at Nordita with the rest of us just Westerners, had insisted that the meeting should be open to everybody." }.  It was organised by Alan Luther, and there were many participants from the Western countries: Alan Luther, Jan Ambjorn,  Bergfinnur Durhuus,  Holger Bech Nielsen, Benny Lautrup, David Gross, Curt Callan, John Kogut, Paolo Di Vecchia, Michael Green, Jean Zinn-Justin  and others, and Lars Brink was one of them.  The Eastern  world was presented by  the colleagues  from the Landau Institute and other Moscow institutions\footnote{ There were series of Landau-Nordita  meetings: "Adsorption and Potts Models" Nordita-Landau Institute Condensed Matter Theory Workshop, Copenhagen, Denmark (1980);   "A Frustrated Spin-Gas Model for Doubly Reentrant Liquid Crystals" Nordita-Landau Institute Condensed Matter Theory Workshop, Göteborg, Sweden (1981); "Frustration and Chaos in Spin-Glasses" Nordita-Landau Institute Condensed Matter Theory Workshop, Copenhagen, Denmark (1984).}. I was invited by Alexander Belavin.  At that time I was working in Yerevan, Fig.\ref{fig2}.

It was an exciting time of fundamental discoveries in the experimental and theoretical high-energy physics: deep inelastic scattering experiments, discovery of charm quark mesons, quantisation and renormalisation of the Yang Mills theory, the discovery of asymptotic freedom\footnote{In 1973 the Princeton preprint of David Gross and Franc Wilczek became available in Yerevan. I was an undergraduate student and was studying the article in great details. What was also especially attractive to me was the title of the article. I was thinking,  "it is not about freedom, but at least it provides an asymptotic freedom!"}, formulation of Quantum Chromodynamics, the topological properties of Yang Mills vacuum fields,  the discovery  of 't Hooft - Polyakov monopole,  the chromomagnetic gluon condensation,   Electroweak and Grand Unification Theories,  the supersymmetric extension of the Poincar\'e group,  the development of the supersymmetric Yang Mills theory and String Theories  \cite{Brink:1976sc}, the fields  in which Lars Brink played a pioneering role \cite{Brink:1976bc,  Green:1982sw, Brink:1982wv, Brink:1981nb}.  Alexander Polyakov presented his new approach to the sting theory quantisation.  Benny Lautrup presented the lattice regularisation and Monte Carlo simulations of QCD. 

There was a cultural program as well, an outdoor party in the forest with a fireplace and barbecue. Lars Brink and Alexander Polyakov went for running in the forest and came back after a long run. It was known within colleagues that Alexander goes for jogging at any of the research time breaks.  What became known to me through  the subsequent meetings and friendship with Lars was that he also would go for jogging at the first opportunity and no matter in which part of the world he was!  At the end of the meeting we were invited to a party at Alexander Polyakov's home, where there were many discussions of new horizons in high-energy  particle physics and we were served by an exceptional French wine.
 
This  "marvellous meeting", as Lars quoted in his article \cite{doi:10.1142/5025},  generated a long-lasting friendship, provided a background for the future collaborations and, in particular, my own collaboration with Lars on a supersymmetric high-spin extension of the Poincar\'e group \cite{Antoniadis:2011re}.   There were many subsequent meetings with Lars after the 1981 Meeting - in Copenhagen, in G\"oteborg when he  invited me to visit the high-energy group at the Chalmers University of Technology, in Miami, where Lars had very friendly relations with the members of the high-energy group of the University of Florida, at the Corfu meeting, Fig.\ref{fig1}, and at the Singapore meeting devoted to the celebration of 60 years of the Yang Mills theory organised by Lars Brink and  Kok Khoo Phua \cite{doi:10.1142/9829}.   
 
It took years before our interests started to converge.  The subject that drove the "convergence" of our interests was Lars' permanent attention and his interest in developing the Supersymmetry and Yang Mills theory  \cite{Brink:1976bc, Green:1982sw, Brink:1982wv},  String Theory \cite{Brink:1976sc} and High Spin Field Theory \cite{Bengtsson:1983pd, Bengtsson:1983pg} \footnote{ In \cite{doi:10.1142/5025} Lars expressed his research interests in that years "...I mentioned to Berezin that I was working on spinning particles and strings."  See also Lars articles devoted to his colleagues, collaborators  and friends \cite{Brink:2015ust, Brink:2017ycp, Berkovits:2017wnn, Brink:2020fie, Brink:2020ssu,  Brink:2022bwn}.}. I think that the reason for his interest in High Spin Field Theory  was the expectation that at the end of the day the string theory should be formulated as a Lagrangian field theory of infinitely many interacting high-spin fields.  It seems to me that explicitly that program of building the string field theory for fundamental interactions was formulated by Witten in \cite{Witten:1988sy} :  "So the central task was to  
\begin{romanlist}[(iv)]
\item find the right degrees of freedom;
\item find the right invariance group; and
\item formulate the right Lagrangian."
\end{romanlist}
 The development of the string theory Lars described  in his article \cite{Brink:2015ust}: "We now had a ghost-free model with both bosons and fermions but with massless vector particles and also a sector with massless particles with spin-2. Later they were called the open and the closed string. This was a great achievement but it was really useless for the purpose it had been invented, namely to describe strong interaction amplitudes and the interest faded."  The dual model of hadrons transforms into a string theory of fundamental interactions. 
 
But the question still remains: Is there a QCD string theory for hadrons? That was the initial point of my own interest  in working on a QCD string. The QCD string should be asymptotically free at short distances, that is, should have the perimeter law behaviour at short distances, (a synonym of) a tensionless string, to generate a confining force at large distances and to be ghost-free in the four-dimensional space time.  The idea was to have a string action which is proportional to the perimeter or to the linear size of the string world-sheet surface instead of its area  \cite{Savvidy:2003dv} : 
\begin{equation}\label{gonihedric}
S= m \cdot L=m \int  \sqrt{(\bigtriangleup(g) X_{\mu})^2} \sqrt{g} d^2 \xi ~~~~\leftrightarrow ~~~~m \int ds
\end{equation}
It is a tensionless string at the classical level, and it was demonstrated  that a string tension can be generated through the quantum-mechanical fluctuations \cite{Savvidy:1993qq}. The quantisation of the model and the calculation of its particle spectrum demonstrated that the whole spectrum consisted of  massless particles of increasing integer helicities $h=\pm1, \pm 2, ..$.  The critical dimension was  $d_c=13$, still far from four dimensions\footnote{In the supersymmetric version of the model  (\ref{gonihedric}) the $d_c=6$ \cite{Savvidy:2002tb} .}.  And again, the model "refused" to be a QCD string, but instead represented a  tensionless string.   The massless states showed an extended high-spin Abelian gauge symmetry that collectively transforms all high-spin fields between  themselves (see formula (64) in \cite{Savvidy:2003fx}).   It was appealing to extend this Abelian high-spin gauge symmetry to a non-Abelian case with an expectation that it may help to formulate the desirable string field theory without referring to a world-sheet geometry \cite{Savvidy:2005fi, Savvidy:2010vb, Savvidy:2015jgv}. 
 
The efforts in this direction finally allowed to formulate the high-spin extension of the Poincar\'e group and of the Lagrangian that contains infinitely many interacting  Yang Mills tensor gauge fields  \cite{Savvidy:2005fi, Savvidy:2010vb, Savvidy:2015jgv}. The question of the main concern  was the quantum-mechanical consistency of this {\it high-spin Lagrangian field theory which has only dimensionless coupling constants} \cite{Savvidy:2005fi, Savvidy:2010vb, Savvidy:2015jgv}. The difficulty was and remains to handle the Lagrangian field theory of infinitely many interacting high-spin gauge fields.   The main question to be answered was:  Is it providing a consistent interaction of high-spin gauge fields in four dimensions?

In collaboration  with  Ignatios Antoniadis  at CERN we were interested to find out if {\it there is a subset  of  interaction vertices in string theory which have dimensionless coupling constants and compare them with the high-spin vertices of  Yang Mills tensor gauge fields}   \cite{Savvidy:2008ks, Antoniadis:2009rd}\footnote{The lower spin interaction vertices in string theory  coincide with the Yang Mills and graviton interaction vertices \cite{Neveu:1971mu, Yoneya:1974jg}. }.  At this point we started an intensive communication with  Lars discussing the structure of vertices investigated by Lars and his collaborators at  the Chalmers University of Technology \cite{Bengtsson:1983pd, Bengtsson:1983pg}. Their early results for the cubic interaction vertices for the massless particles gave a hope that a consistent high-spin field theory should exist. Their calculations were made in a light-cone frame that operates  with the physical helicity states.  There was consistency between  vertices in both theories   \cite{ Savvidy:2008ks, Antoniadis:2009rd,Savvidy:2010bk}. 

Lars was saying  that  "the calculations are simplified and transparent on the light-cone \cite{Brink:1982pd, Brink:2017ycp}"  and that "good things should be supersymmetric" \cite{Brink:1976bc, Brink:1976sc, Green:1982sw, Brink:1982wv, Brink:1981nb}.  Lars was interested if there exists a supersymmetric extension of the Lagrangian proposed in  our earlier publications \cite{Savvidy:2005fi, Savvidy:2010vb, Savvidy:2015jgv} or at least the supersymmetric extension of high-spin Poincar\'e algebra \cite{Savvidy:2008zy, Savvidy:2010vb, Savvidy:2010kw} .  In fall of 2010 we arranged a meeting  during the conference in Fort Lauderdale (Coral Gables) of Florida organised by Thomas Curtright. The aim was  to investigate a possible supersymmetric extension of the high-spin Poincar\'e algebra by adding high-spin anti-commuting generators.  Initially  we were always  getting the inconsistent algebras (see footnote \ref{fut1}).  As always, Lars was jogging in the morning hours  in the beautiful  sandy beaches of Fort Lauderdale.   At some moment we tried the most simple structure, and that was working! I returned to CERN and it took few months to formulate the results and investigate the irreducible representations. The  supersymmetric generalisations of high-spin de Sitter and conformal groups  were also suggested in our publication \cite{Antoniadis:2011re} . 

There was an intensive exchange of messages with Lars concerning the concept of {\it tensionless strings, conformal symmetry  and continuous spin representations}.  Few messages from Lars that discussed these problems  are reproduced below:   On February 15, 2011."Dear George, It looks very good. I will probably not have too much time this week since I have to do Nobel work and next week I will be in India. After that I will spend one month at IAS in Princeton so then I will [have] lots of time. I am as always peeking. I am thinking about extended susy. I see no problems with that. However, the representations look a bit complicated since you will  [have] a degeneracy on each level. That could in fact be quite interesting. Best wishes, Lars."  On February 18, 2011.  "Dear friends, This is a very clever and interesting argument. Pierre and I tried to find the continuous spin reps in the conformal algebra but we could not do it, which made us quite unhappy at the time. It would be good (or in fact bad) if we convince ourselves that we do not have conformal invariance in the generalised gauge theories. I do not remember our argument in detail but we did go through the conformal algebra and at some stage we could not keep the $T^i$'s in the $J^-$. I have been solidly busy the whole week  [working in Nobel Committee]. Next week I will be in India and from March 3 I will be at IAS for a month with hopefully oceans of time  \footnote{
On March 8, 2011.”I agree, but I feel like a calf in the spring when it gets out in the field. (This is a Swedish saying. In Greece they might be out all year around.) It is so nice to be in a place where I can concentrate on research.
The Kalb-Ramond field is conformal in six dimensions if you couple it in the (2,0) supermultiplet, which shows that it is not conformal by itself. I’ll go over the manuscript this afternoon. Best wishes, Lars ”}."

On February 27, 2011. "Dear Lars, concerning the interesting question of conformal invariance.  .... We calculated ... the trace of high-spin energy momentum tensor in order to get convinced that the theory is conformally invariant and it is. At the same time I think that I understand your result about the absence of the infinite representations in the case of conformal group, I will try to elaborate on that in the next messages.  Where are you, in India? Should be more sunny than here! All the best, George".  On February 21, 2011.  "Hi,  it seems that the group can be extended to the conformal case as well, but I don't know about representations yet, it tells interesting things about "gauge generators".  The superconformal extension is yet to be studied,  the updated file contains more details. All the best, George".  

On February 27, 2011. "Dear George, I came back last night from Pune in India. A former student of Pierre, Sudarshan Ananth, whom I have worked  lot with too has set up a new group there and he organised a meeting with some forty people. It was very nice and very interesting. The most interesting thing was to see how good the undergraduate students are. For the top school they choose between the 1 percent  best students in the country!  Asia is really taking over. I will now be home for three days before going to Princeton. Two of those days I will spend in Stockholm [working in Nobel Committee] and tomorrow I will prepare everything for the trip. I won't have time to really read carefully until I am installed in Princeton but then I will have a lot of time. Best wishes, Lars". 

On March 7, 2011.   "Dear George, ...Did you ever try to have L's with spinor indices instead of vector indices? In principle you can rewrite every vector index $\mu$ with two-component spinors $\alpha \dot \beta$. I have to think about it whether it really generalises anything. Best, Lars". 

On March 8, 2011.  "Dear Lars, I was thinking almost in the same way as you! My path is as follows: we have $P^{\mu}$ vector in the spacetime algebra and what we do? We ask if one can take the square root of $P^{\mu}$ and arrive to the supercharges. Now we have many $L$ tensors  so we can ask why we should not take the square root of $L$? ... I did some calculations in this direction, but I am just running out of steam. Let us finish with this and go to the next one: either with conformal questions or large supercharges. ... All the best, George".  On March 9, 2011. "Dear George and Ignatios, I have now gone through it once more and added in some text. ...I have reshuffled the references and I do not see any citation to Fierz in the text but it is a beautiful citation so I hope it has a place somewhere. As I wrote to George I am not fully satisfied with the text after (28)   ... Otherwise I like it a lot. Best wishes, Lars" . 

On March 15, 2011. "My problem with identifying this to the zero-tension limit is the multiplicities with the higher-spin states. How many survive in the limit? It cannot be an infinity as one might believe by just rotating all Regge trajectories. I have thought about this problem of and on for 35 years! ..."  

On February 15, 2011. "Dear Lars, attached is the draft file. The question of conformal symmetry is very interesting and  will be nice to think about that. By the way Ignatios in his last article discusses the question of relation between conformal and scaling symmetries. Why there is no continuous rep. in conformal group? What I know is that the group can be extended into the AdS space background. May be it is worth to consider?  Have a nice trip to India! All the best, George" 

On  February 18, 2011. "Dear George, A comment concerning the question of Lars on conformal symmetry. It is not necessary that a massless theory is conformal. In my recent paper with Buican that you quote, we had proved that scale implies conformal invariance under some assumptions: unitarity, supersymmetry and existence of a well defined dilatation current. .... I suspect that massless higher spin theories in 4 dims may have similar behaviour but it requires separate analysis. Best regards, Ignatios".

\begin{figure}
\centerline{  \includegraphics[width=2.5cm]{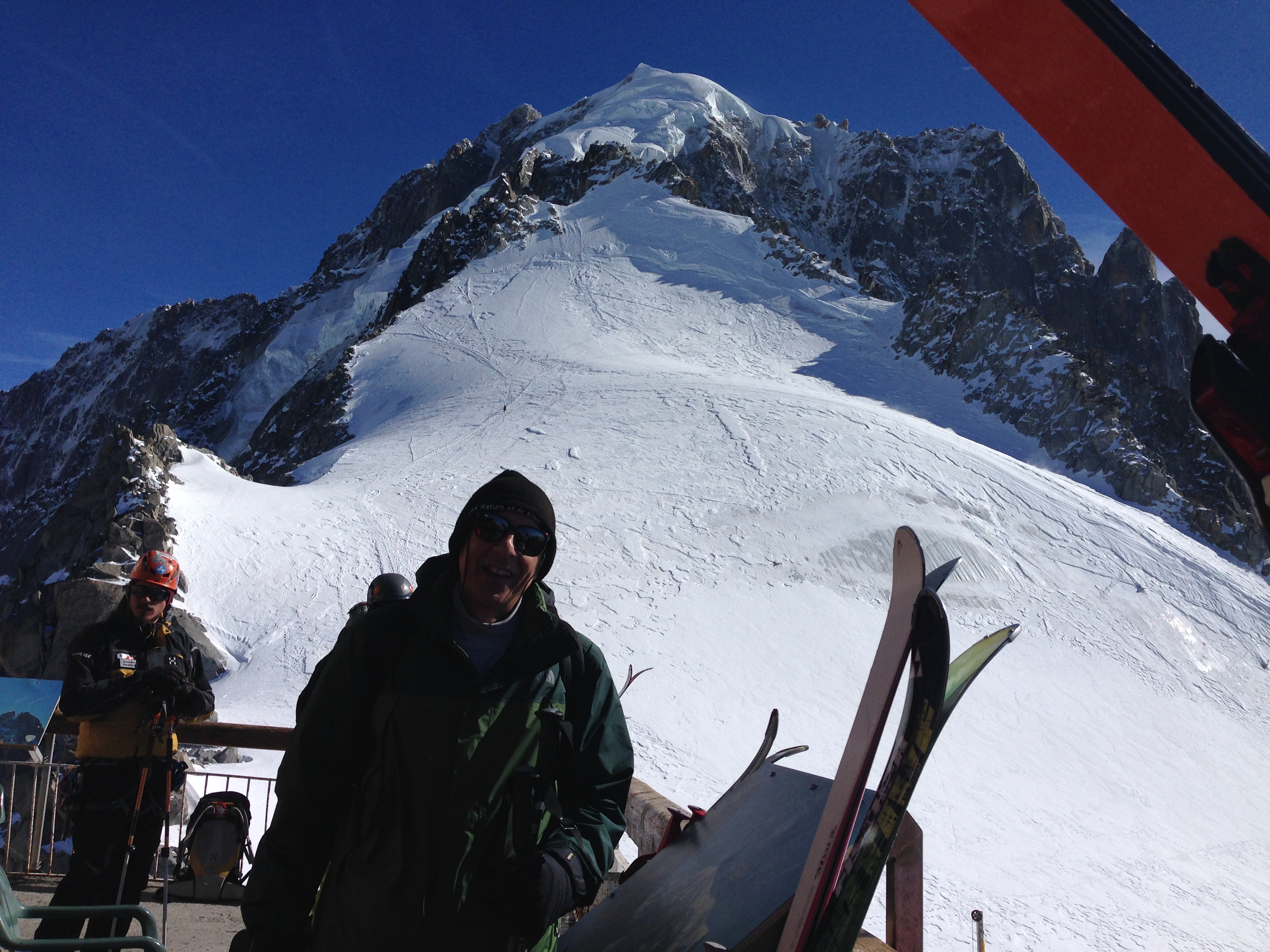}~~~~~
  \includegraphics[width=2.5cm]{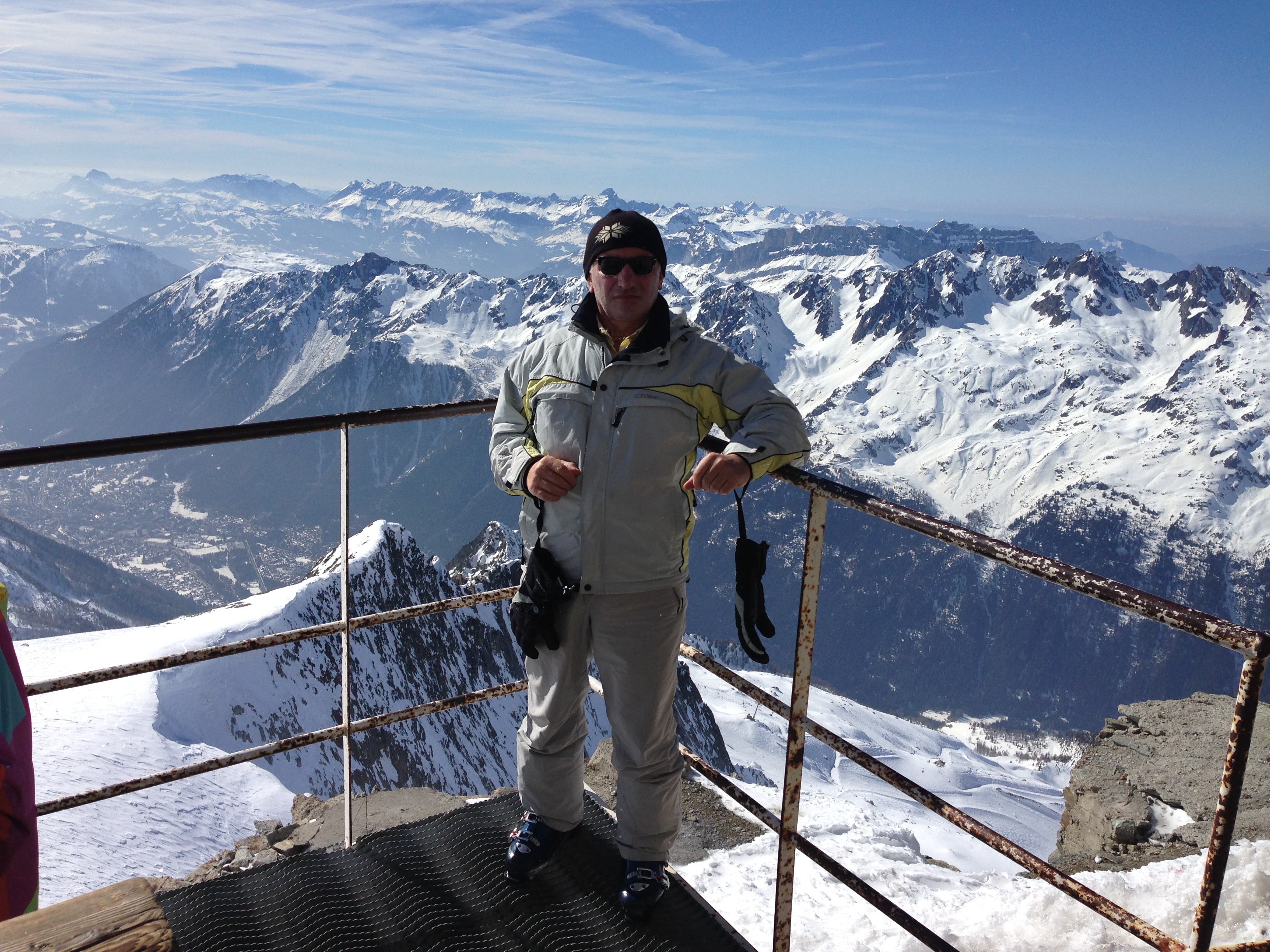}}~~~~
\caption{ It was great to ski with Ignatios, adventure on French-Swiss-Italian Alps, all black pistes ski and magnificent off-pistes as well.  Photos are at the top of Argenti\'ere glacier before descending.} 
\label{fig3}
\end{figure}

I told to Lars that with Ignatios we were going to ski on the weekend Fig. \ref{fig3}. Lars responded, "Have a nice skiing weekend. Where do you go these days, Avoriaz, Flaine, La Cluzaz, Moriond.... I do envy you. We have had ice for the last six weeks. Best wishes, Lars". Unfortunately I did not have a chance to ski with Lars in Swiss-French  Alps to witness not only how good he is in jogging,  but also how perfect he is in skiing.  

 \section{Acknowledgments}
 I would like to thank Professor  Kok Khoo Phua from the Institute of Advanced Studies (IAS) at Nanyang Technological University  for organising the meeting "Remembering Lars Brink".  My thanks are also to Jan Ambjorn, Bergfinnur  Durhuus and Yuri Makenko for their recollections concerning the Landau - Nordita meeting held in 1981 and  for a number of suggestions which have improved the presentation.

\bibliographystyle{ws-rv-van}
\bibliography{ws-rv-savvidy}


\end{document}